\documentclass[preprint,preprintnumbers,amsmath,amssymb]{revtex4}
\usepackage{graphicx}

\begin{document}

\title{Kinetics of proton pumping in cytochrome c oxidase}

\author{ Anatoly  Yu. Smirnov$^{1,2,3}$, Lev G. Mourokh$^{4}$, and Franco Nori$^{1,3}$}

\affiliation{ $^1$ Advanced Science  Institute, The Institute of Physical and
Chemical Research (RIKEN), \\
Wako-shi, Saitama, 351-0198, Japan \\
$^2$ CREST, Japan Science and Technology Agency, Kawaguchi, Saitama 332-0012, Japan \\
$^3$ Center for Theoretical Physics, Physics Department, The University of Michigan, Ann Arbor, MI 48109-1040, USA,\\
$^4$ Department of Physics, Queens College, The City University of New York, Flushing, New York 11367, USA}

\date{\today}

\begin{abstract}
We propose a simple model of cytochrome c oxidase, including four redox centers and four protonable sites, to study the time
evolution of electrostatically coupled electron and proton transfers initiated by the injection of a single electron into the
enzyme. We derive a system of master equations for electron and proton state probabilities and show that an efficient pumping of
protons across the membrane can be obtained for a reasonable set of parameters. All four experimentally observed kinetic phases
appear naturally from our model. We also calculate the dependence of the pumping efficiency on the transmembrane voltage at
different temperatures and discuss a possible mechanism of the redox-driven proton translocation.
\end{abstract}


\maketitle

\section{Introduction}

The last enzyme of the respiratory chain of animal cells and bacteria, cytochrome $c$ oxidase (CcO), operates as an efficient
nanoscale machine converting electron energy into a transmembrane proton electrochemical gradient
\cite{Alberts02,Nicholls92,Wik77,Wik04,Gennis04,Brez04,Branden06}. The ATP (adenosine triphosphate) synthase enzyme uses  this
energy to synthesize ATP molecules serving as the ``energy currency" of the cell. The process of energy conversion starts when a
molecular shuttle, cytochrome $c$, delivers, one by one, high-energy electrons to a dinuclear copper center, Cu$_A$, located near a
positive side (P$-$side) of the inner mitochondrial membrane (see Fig.~1). In recent time-resolved optical and electrometric studies
\cite{BelPNAS07} of the CcO transition from the oxidized (O) state to the one-electron reduced form (E), a \emph{single} electron is
donated to the Cu$_A$ redox center by a laser-activated molecule of ruthenium bispyridyl (RubiPy). Thereafter, in a few microseconds
($\sim$10 $\mu$s), a major part of an electron density ($\sim$70\%) is transferred from the Cu$_A$ center to the low-spin heme $a$
(Fe-$a$). Heme $a$ is located within the membrane domain at a distance about 2/3 of the membrane width, $W$, counting from the
N-side \cite{BelPNAS07,WV07,MedStuch05}. Within a time interval of approximately 150~$\mu$s, about 60\% of the electron population
is transferred from heme $a$ to heme $a_3$ (Fe-$a_3$). Heme $a_3$, jointly with the next electron acceptor in line, a copper ion
Cu$_B$, form a binuclear center (BNC, sites $R$ and $B$ in Fig.~1), serving as an active catalytic site for dioxygen reduction to
water. The redox centers $a,$ $a_3,$ and Cu$_B$, are located approximately at the same distance $(2/3~W)$ from the N-side of the
membrane as heme $a$.

The next phase (with a time scale of the order of 800~$\mu$s) is characterized by a complete electron transfer to the copper ion
Cu$_B$. Time-resolved measurements \cite{BelPNAS07} show that the first ``10~$\mu$s" phase of the electron transfer process is not
accompanied by a proton transfer, but the ``slowness" of the second ``150~$\mu$s" and the third ``800~$\mu$s" phases hints to the
proton participation during phases.

The proton path from the negative side of the membrane (N-side) toward the P-side (for pumped protons) and toward the binuclear
center (for substrate or ``chemical" protons) goes through the residue $E278$ (for the $Paracoccus~ denitrificans$ enzyme
\cite{BelPNAS07}). These residues are located at the end of the so-called D-pathway (Fig.~1). A fraction of the substrate protons
can also be delivered to the BNC via an additional K-pathway, which we will not consider here. The proton to be pumped is supposed
to move from $E278$ (schematically shown as the site ``$D$" in Fig~1)  to an unknown protonable ``pumping" site $X$ (likely a heme
$a_3$ propionate), located above the BNC \cite{BelNat06}, and, thereafter, via an additional protonable site $C$
\cite{Siletsky04,Brez06}, to the P-side of the membrane. After a fast reprotonation from the D-channel, the residue $E278$ can
donate a substrate proton to the catalytic site near the BNC (probably, to an OH$^-$ ligand of Cu$_B$ \cite{BelPNAS07,WV07}). It is
assumed \cite{BelPNAS07,WV07} that during the second ``150~$\mu$s" phase, the first (pre-pumped) proton moves from the residue
$E278$ (the site ``D" in Fig.~1) to the pump site $X$, whereas in the third ``800~$\mu$s" phase, the second (substrate or chemical)
proton populates a catalytic site $Z$ near the BNC. In the final phase, which occurs in 2.6 ms, the first proton (in $X$) is
translocated (via $C$) to the P-side, which is characterized by a higher electrochemical potential than the N-side of the membrane.

As a result of all these processes, two protons are taken from the N-side of the membrane, and one electron is taken from the
P-side, and eventually one proton is pumped to the P-side. Moreover, one proton and one electron are consumed at the catalytic site
to finally produce a water molecule around the BNC. It should be noted that kinetic phases with similar time scales
($10~\mu$s$~\rightarrow 100~\mu$s$~\rightarrow 1000~\mu$s) have been revealed in other CcO enzymes at various transition steps
between the states of the enzyme \cite{Ruit02,Siletsky04, SalomBrez05}.

Kinetic data obtained in experiments \cite{BelPNAS07,Siletsky04,Ruit02,SalomBrez05} reflect important details of the still elusive
proton pumping mechanism in cytochrome $c$ oxidase. To extract these details and gain a deeper insight into the operating principles
of the CcO proton pump, it is necessary to compare results of experiments with theoretical predictions. In Ref.~\cite{Olsson06}, a
simplified empirical valence bond (EVB) effective potential was combined with a modified Marcus equation to model time-dependent
electron and proton transfers in CcO in the range of milliseconds. However, this approach was applied to the single transfer event,
not to the sequence of events, and the obtained time scale (one microsecond) differs by orders of magnitude from the experimental
data (about 100 microseconds).
 A computational analysis of the CcO energetics was presented in Refs.~\cite{Olsson07,Pisliakov08,SiegBlom07,SiegBlom08,Quenneville06}
  with molecular models reproducing energetic barriers for the proton transfer steps \cite{Olsson07,Pisliakov08}. The obtained
energetic map of the proton and electron pathways in the CcO enzyme can be converted into a set of rate constants, which
qualitatively explains the kinetics and unidirectionality of the pumping process. However, these studies do not result in a
quantitative model of the efficient CcO proton pump. Moreover, the error range of these semi-microscopic calculations ($\sim$~2
kcal/mol) is sometimes higher than the difference between the energy barriers \cite{Pisliakov08,SiegBlom08}.

Kinetic models of the proton pumping process were also discussed in Ref.~\cite{KimPNAS07}. Within the master equation approach, it
was shown that the proton pumping effect can be achieved in a simplified system having one redox and two proton sites and, with a
higher efficiency, $\eta \sim$~0.9, for the design with two redox and two protonable sites, which are electrostatically coupled to
each other. However, this work does not contain any predictions for the kinetics of the pumping process in more realistic set-ups,
with at least four redox sites (Cu$_A$, heme $a$, heme $a_3$, and Cu$_B$) and two protonable sites (a residue $E278$ and a pump site
$X$). To find proper parameters for the proton pump, the authors of Ref.~\cite{KimPNAS07} resort to a Monte Carlo search in a
multidimensional parameter space. It is hard to imagine, however, that a random search can provide a reasonable set of parameters
which will comply with all physical restrictions of real pumps. In general, for comprehensive theoretical studies, it is preferable
to determine the relevant parameters of the system using detailed microscopic calculations (see, e.g.,
Refs.~\cite{Olsson06,Olsson07,Pisliakov08}). However, the huge computational complexity of biological structures makes such an
approach extremely difficult. In our paper, we include reasonable estimates for the system parameters into a model describing almost
simultaneous electron and proton transfer processes and compare the obtained kinetics to experimentally observed time scales and
site populations of cytochrome c oxidase \cite{BelPNAS07}.

The time evolution of the proton pumping process in CcO, related to the experimental data of
Refs.~\cite{BelPNAS07,Ruit02,Siletsky04,SalomBrez05}, was discussed in
Refs.~\cite{WV07,MedStuch05,SiegBlom07,SugStuch08,SiegBlom08}. In these works, the kinetics of the electron-proton system is broken
down into a cascade of quasi-equilibrium states characterized by distributions of electrons and protons over the sites, as well as
by a set of transition rates corresponding to specific kinetic phases. It should be emphasized, however, that many electron and
proton transfers can be separated by only a nanosecond time scale, and, consequently, the experimentally observed kinetic rates
comprise contributions of several almost-simultaneous individual electron and proton transfer events \cite{MedStuch05,SugStuch08}.
Correspondingly, an approach taking into account the kinetic inseparability of electron and proton transitions can be useful for
understanding recent experimental findings \cite{BelPNAS07}. We note that kinetic coefficients used in the theoretical analysis of
Refs.~\cite{MedStuch05,SiegBlom07,SugStuch08,SiegBlom08} were deduced from experiments without independent microscopic calculations
of the heights of individual electron and proton barriers.

In the present paper, we analyze electron and proton kinetics in cytochrome $c$ oxidase within a simple physical model including
four redox centers and four protonable sites electrostatically coupled to each other in the presence of a dissipative environment.
Using the master equation approach, we reproduce all four kinetic phases observed in Ref.~\cite{BelPNAS07} for a reasonable set of
parameters. It should be emphasized that we have performed extensive numerical studies for a wide range of parameters and we found
that our model of proton pumping is quite robust to significant variations of the parameters. The specific set of parameters
presented below gives a very good agreement with the experimental data of Ref.~\cite{BelPNAS07}. We consider a single cycle of
events, which starts at $t = 0$ with one electron transfer to the Cu$_A$ center and finishes at the moment $t = t_B$, when the redox
site Cu$_B$ is completely reduced. Notice that the injection of additional high-energy electrons is necessary to maintain this
nonequilibrium state of the CcO enzyme. We also determine the efficiency of the proton pumping for our model and its dependencies on
the temperature and transmembrane voltage.

 The rest of paper is structured as follows. Our model and
its parameters are presented in Section II. Results of numerical studies are shown in Section III and discussed in Section IV.
Section V contains the conclusions of our work. The detailed derivation of the master equations and the measurable variables is
presented in the Appendix. It should be noted that while the results of this paper are obtained in the \emph{classical} regime, our
approach (based on quantum transport theory) can be used to examine fine \emph{quantum} effects and, consequently, the detailed
derivation is worth presenting here.

\section{Model}

As in the real CcO enzyme \cite{Iwata95,Tsukihara96,Yoshikawa98,Yoshikawa06,Ostermeier97}, the redox chain of the present model
includes four centers: Cu$_A$ (site $A$), heme $a$ (Fe-a, site $L$), heme $a_3$ (Fe-$a_3$, site $R$), and Cu$_B$ (site $B$), as
schematically shown in Fig.~1. The transport chain for protons has four sites: $D$ (presumably related to the residue $E278$ near
the end of the D-pathway), $X$ (the pump site above the BNC), a protonable site $C$ placed on the way from the $X$-site to the
P-side of the membrane, and, finally, a protonable site $Z$ located in the proximity of the BNC and related to the OH$^-$ ligand of
Cu$_B$ (see Fig.~1). The sites $B$ and $Z$ serve as final destinations for the injected electron and for the substrate proton,
respectively.  We assume that the electron can be transferred between the pairs of redox states $A$ and $L$, $L$ and $R$, $R$ and
$B$; and that protons can be translocated between the pairs of protonable sites $D$ and $X$, $X$ and $C$, as well as $D$ and $Z$.

To provide an ``openness" of the CcO enzyme, which is inherent in the living systems \cite{KimPNAS07}, we allow proton transitions
between the site $D$ and the negative side of the membrane as well as between the site $C$ and the positive side of the membrane.
Protons are delivered to the  catalytic site $Z$ partially through the K-pathway \cite{Gennis04,WV07}. This channel can be
incorporated into our model, but, for simplicity, it will be neglected. The N- and P- sides of the membrane play roles of proton
reservoirs which work as a source (N-side) and a sink (P-side) of protons for the enzyme. The redox sites are disconnected from
electron reservoirs, and \emph{only one} electron is injected into the redox chain at the initial moment of time, $t=0$.

 With the condition of single-occupation of each individual site, the system can
be populated with up to four protons.  Following the setup of Ref.~\cite{BelPNAS07}, we assume that CcO is populated with a single
electron initially located on site $A$. To quantitatively describe this system we introduce 64 basis states $|\mu\rangle,\, \mu =
1,\ldots,64$ (see Appendix). The time evolution of the probability distribution over the basis states, $\langle \rho (t)\rangle  =
\{\langle \rho_{\mu} (t) \rangle\}$, is governed by the system of master equations, Eq.~(\ref{RhoMFin}), with the solution given by
Eq.~(\ref{RhoTime}) in the Appendix. The time-dependent probability distribution $\langle \rho(t)\rangle$ allows us to determine the
average populations of all electron and proton sites, $\langle n_{\alpha}\rangle$ and $\langle N_{\beta}\rangle$, as functions of
time. We can also calculate the number of protons, $\langle N_{\rm P}(t) \rangle$, translocated to the positive side of the membrane
[see Appendix, Eq.~(\ref{Nsigma})]. The value of $\langle N_{\rm P} \rangle$ taken at the end of the pumping cycle ($t = t_B$)
determines the pumping efficiency $\eta$ defined \cite{KimPNAS07} as the number of protons pumped across the membrane per electron
consumed:
\begin{equation}
\eta = \langle N_{\rm P}(t_B) \rangle. \label{eta}
\end{equation}
Note that the efficiency $\eta$ can take negative values in the case when protons move back from the positive side to the negative
side of the membrane.

\subsection{Electrostatic interaction}
The electrostatic interaction between the redox ($\alpha = A,L,R,B$) and protonable ($\beta = D,X,C,Z$) sites plays a pivotal role
in the electron-proton energy exchange. It should be noted that we consider here only \textit{direct} Coulomb interactions between
electron and proton subsystems and between protons themselves. This removes strict geometrical restrictions on the relative
positions of electron and proton active sites imposed in our previous model \cite{PumpPRE08} based on the F\"orster-type energy
exchange between electrons and protons. Microscopic calculations of the electrostatic parameters, $u_{\alpha\beta}$ and
$u_{\beta\beta'}$, involved in the Hamiltonian $H_0$ [see Appendix, Eq.~(\ref{H0})], require a detailed knowledge of the CcO
structure complemented by the comprehensive dielectric map of the enzyme \cite{Olsson07,Warshel06,Schutz01}. Instead, we tune the
Coulomb energies to get the best possible fitting of the time scales and site populations measured in the experiment
\cite{BelPNAS07}. The obtained values of Coulomb parameters correlate well with information about the distances between the active
sites \cite{Iwata95,Tsukihara96,Yoshikawa98,Yoshikawa06,Ostermeier97} for reasonable values of the effective dielectric constants.

To describe the experimentally observed kinetic phases of the pumping process, we assume that the coupling, $u_{BZ} = 630$ meV,
between the copper ion Cu$_B$ and the catalytic site $Z$ (likely an OH$^{-}$ ligand of Cu$_B$ \cite{BelPNAS07,WV07}) and the
coupling, $u_{RX}=555$ meV, between heme $a_3$ and the pump site $X$ are higher than the electrostatic energies $u_{RZ}=530$ meV and
$u_{BX}=u_{XZ}=510$ meV. Structural studies of the CcO enzyme \cite{Iwata95,Tsukihara96,Yoshikawa98,Yoshikawa06,Ostermeier97}
performed at a resolution of about 2 \AA\ show that the BNC redox sites $R$ (heme $a_3$), $B$ (Cu$_B$) and the protonable sites $X$
and $Z$ are separated by a distance of the order of 6 \AA. The value of the electrostatic coupling between these sites, $u \sim 600$
meV, roughly corresponds to the effective dielectric constant, $\epsilon \sim 4$, which is of frequent use for a description of a
dry protein interior \cite{Olsson07,SiegBlom08,Quenneville06}. It should be emphasized, however, that the concept of dielectric
constant is not completely appropriate for a calculation of Coulomb potentials in the heterogeneous environment inside and near the
BNC \cite{Warshel06,Schutz01}.

The distances, $r_{LD}, r_{RD}$, between the residue $E278$ (site D) and the sites $L$ and $R$ are almost the same: $r_{LD} = 12.3$
\AA,\, $r_{RD} = 12.8$ \AA\, \cite{Ostermeier97,Michel98}. We estimate the electrostatic coupling between these sites as $u_{LD}
\sim u_{RD} \simeq 75$ meV, which corresponds to the higher dielectric constant $\epsilon \sim 15.$ We consider a smaller dielectric
constant, $\epsilon \sim 10$, for the interaction, $u_{LX} = 100$ meV, between the sites $L$ and $X$ separated by the distance
$r_{LX} \sim 14.2$ \AA\, \cite{SiegBlom08}. Distant-dependent dielectric constants, $\epsilon(r_{\alpha\beta})$, are common in
protein electrostatics \cite{Olsson07,Warshel06,Schutz01}.

Note that here, as in the models of Refs.~\cite{WV07,SiegBlom07}, the electrostatic coupling, $ u_{RX},$ between heme $a_3$ (site
$R$) and the site $X$ is stronger than the interaction, $ u_{LX},$ between heme $a$ (site $L$) and the pump site $X$. For the other
parameters we choose the following values (in meV): $u_{DX}\sim 60,\  u_{DZ}\sim u_{BD}\sim 70,\ u_{XC}\sim 100,\  u_{AD}\sim 25,\
u_{AZ}\sim 20.$ The Coulomb energies $u_{CZ},u_{DC},u_{AX}, u_{AC},u_{RC},u_{LC},$ and $u_{LZ}$ are assumed to be near 30 meV.
Despite the fact that these energies are about or higher than the temperature energy scale, $ T = 298$ K $\sim 26$~meV, they have a
minor influence on the performance of the model.

\subsection{Energy levels of the sites}
We assume that the difference $\Delta \mu$ (\ref{DMu}) between the electrochemical potential $\mu_{\rm P}$ of the P-side  and the
potential $\mu_{\rm N}$ of the N-side of the membrane is about 210 meV  at standard temperature, $T = 298$ K, with $\mu_{\rm P} =
105$ meV and $\mu_{\rm N} = - 105$ meV. This corresponds to voltage $V\simeq 150$ meV applied across the membrane. We include the
electron charge in the parameter $V$ and measure voltage, along with other energies, in units of meV. According to
Eqs.~(\ref{EnergyV}), the energy levels, $\varepsilon_{\alpha}$ and $\varepsilon_{\beta}$, of the electron and proton centers are
shifted from their intrinsic values $\varepsilon_{\alpha}^{(0)}$ and $\varepsilon_{\beta}^{(0)}$ depending on the voltage $V$ and on
the positions $x_{\alpha}, x_{\beta}$ of the active sites. To estimate the electron and proton energies, we take into account the
facts \cite{Brez04} that cytochrome $c$ delivering electrons to the CcO enzyme has a redox potential of order of 250 meV, and that
the total drop of electron energy between cytochrome $c$ and the dioxygen reduction site $B$ is about 550 meV. The equilibrium
midpoint potentials \cite{BelPNAS07,WV07} of the Cu$_A$ center ($E_m \simeq 250$ meV) and heme $a$ ($E_m \simeq 270$ meV) can also
be used as a general guide for estimating energies \cite{Moser06}, although the real parameters can deviate from the estimated
values.

We find that our model performs with the high efficiency, $\eta \sim 0.95$, and reproduces all experimentally observed kinetic
phases \cite{BelPNAS07} for the following set of electron intrinsic energies (in meV): $\varepsilon_A^{(0)} = -175,\
\varepsilon_L^{(0)}=-240,\ \varepsilon_R^{(0)}=-185,\ \varepsilon_B^{(0)}=-155,$ and for the following energies of protonable sites
(in meV): $\varepsilon_D^{(0)}=-100,\ \varepsilon_X^{(0)}=250,\ \varepsilon_C^{(0)}=195,$ and $\varepsilon_Z^{(0)}=-65.$ It should
be noted that in the presence of the transmembrane voltage, $V = 150$ meV, the electron energy levels of $A$ and $L$ sites,
$\varepsilon_A = -250,\ \varepsilon_L = -265,$ are close to the values extracted from equilibrium redox titrations (see also
Ref.~\cite{Farver06}, where an estimation, $(\varepsilon_A - \varepsilon_L) \simeq 18$~meV, has been obtained). For energies of
other redox sites we use the values: $\varepsilon_R = -210,\ \varepsilon_B = -180.$ The energies of the protonable sites are also
shifted with voltage, $V = 150$ meV, present: $\varepsilon_D = -85,\ \varepsilon_X = 295,\ \varepsilon_C = 270,$ and $\varepsilon_Z
= -40$. It should be stressed that the energy, $\varepsilon_X$, of the pump site $X$ is set to be higher than the potentials of the
proton reservoirs on both sides of the membrane: $\varepsilon_X > \mu_{\rm P} > \mu_{\rm N}.$ However, the presence of an electron
on the site $R$ decreases the proton energy to the level, $\varepsilon \sim (\varepsilon_X - u_{RX}) \sim - 260$ meV, which is below
the energy of the $D$-site and below the electrochemical potential, $\mu_{\rm N} = -105$ meV, of the N-side of the membrane. As a
result, the pump site $X$ is populated with a pre-pumped proton. When the chemical proton moves to the site Z and the electron is
transferred to the $B$-site, the energy level of the $X$-site returns to the initial position, $ \varepsilon_X=295$ meV, since the
electron and proton charges of the catalytic site compensate each other, $u_{BX}=u_{XZ}$. The high-energy pre-pumped proton can now
move to the site $C$ and, after that, to the P-side of the membrane characterized by the potential $\mu_{\rm P} = 105$ meV. A large
energy gap, $(\varepsilon_X -\varepsilon_D) \sim 380$ meV, significantly suppresses the return of the $X$-proton to the site $D$ and
to the N-side of the membrane.

\subsection{Reorganization energies and transition rates}
Part of the energy delivered to the redox center Cu$_A$ at the initial time, $t = 0$, is  dissipated to an environment characterized
by sets of electron ($\lambda_{\alpha \alpha'}$) and proton ($\lambda_{\beta\beta'}$) reorganization energies. To be efficient, the
proton pumping process should occur with minimal energy dissipation. It is shown in Ref.~\cite{Jas05} that the reorganization energy
for the $a$ to $a_3$ electron transfer in the CcO enzyme can be as low as 100 meV. Similar estimates apply for the proton
reorganization energies \cite{Silverman00,Wraight05}. Here, we use the higher energy parameter, $\lambda_{AL} = 200$ meV, for the
A-to-L transfer and accept the lower value, $\lambda_{\alpha \alpha'}\simeq \lambda_{\beta\beta'} \simeq 100~$meV, for other
electron and proton transitions. It is argued in Refs.~\cite{Farver06,Larsson95,Ramirez95}, that for the Cu$_A \rightarrow $ heme
$a$ electron transition the reorganization energy must be in the range from 150 meV to 500 meV, which is much lower than the typical
values of the reorganization energy for electron transfers in protein. The low values of electron reorganization energies ($\lambda
\sim 2$ -- 4~kcal/mol) have also been calculated for electron transfer reactions in \textit{Rhodobacter sphaeroides}
\cite{Warshel06}.

To reproduce the initial kinetic phases, we use the following tunneling energies: $\Delta_{AL} \sim 0.9\ \mu$eV, and $\Delta_{LR}
\sim \Delta_{BR} \sim 14\ \mu$eV. The parameters $\Delta_{LR}$ and $\Delta_{BR}$ describe the electron transfers, which are  coupled
to the slower proton transitions characterized by the energy scales: $\Delta_{DX} \sim \Delta_{CX} \sim 0.3\ \mu$eV, and
$\Delta_{ZD} \sim 0.2\ \mu$eV. It should be noted that the electron transfer between heme $a$ and heme $a_3$ can occur in a
nanosecond time scale \cite{Pilet04}. The hydrogen-bonded chains in proteins are also able to conduct protons in nanoseconds or
faster \cite{Nagle78,Zundel95}.

We also select the values $\Gamma_{\rm N} \sim \Gamma_{\rm P} \sim 17$ ms$^{-1}$ for the parameters $\Gamma_{\rm N}$ and
$\Gamma_{\rm P}$, which determine the flow of protons through the enzyme. These parameters $\Gamma_{\rm N}$ and $\Gamma_{\rm P}$ are
of the same order as some of the transition rates $\kappa_{\mu\nu}$ used in Ref.~\cite{KimPNAS07}.

\section{Results}

\subsection{Four kinetic phases}
In Fig.~2, starting at $t = 0.1~\mu s$, we show a process of population and depopulation of the electron, $ n_{A,L,R,B} $, and
proton, $ N_{D,X,C,Z}$, sites as well as the time dependence of the average number of protons pumped to the positive side of the
membrane, $ N_{\rm P}$. From here on we drop the brackets $\langle \ldots\rangle$ denoting the averaging over the environmental
fluctuations and over the states of the proton reservoirs. The calculations are performed for the standard conditions ($\mu_{\rm P}
= 105$ meV, $\mu_{\rm N} = -105$ meV, $\Delta pH = -1$, $T = 298~K$) and for the transmembrane voltage $V = 150$~meV. We assume that
initially a single electron is located at the site $A$ (Cu$_A$), and a proton occupies the site $D$. This means that at $t = 0$ only
one element of the density matrix is not equal to zero: $\rho_2 (0) = 1.$

During the first phase of the process the electron moves from the site $A$ to the site $L$ (heme~$a$). In $\sim$~10 $\mu$s near 70\%
of the electron density is transferred to the heme $a$ (site $L$) with the remaining 30 percent distributed almost equally between
the site $A$ (Cu$_A$) and the site $R$ (heme~$a_3$). This corresponds roughly to the 70 percent electron population of heme $a$
after the first $10~\mu$s phase observed experimentally in Ref.~\cite{BelPNAS07}. No pronounced changes in populations of the
protonable sites accompany this stage [see Fig.~2~(b)].

The second phase of the electron transfer is postponed by the time 150 $\mu$s, despite the fast intrinsic transition rate between
the $L$ and $R$ redox sites. Besides the 55 meV potential difference between the sites $R$ and $L$, the electron transfer in this
phase is hampered by the involvement of protons. It is evident from Figs. 2a and 2b that, with a microsecond delay, the slightly
uphill electron transfer from the site $L$ to the site $R$ is followed by the proton translocation from the site $D$ ($E_D=-85$ meV)
to the pump site $X$ having much higher initial energy, $E_X=295$ meV. This transition has been made possible by the strong $R$-$X$
Coulomb attraction ($u_{RX} $~=~555 meV) lowering the effective energies of both electron  and proton sites. In line with the
experimental data \cite{BelPNAS07} at $t = 150~\mu$s, the electron density is located mainly on the site $R$ (60\%) and partially on
the sites $L$ ($\sim$20\%), and on the site $B$ ($\sim$15\%). The site $A$ is practically empty at this stage. It is important that
at almost the same moment of time ($t \approx 150~\mu$s) the population of the protonable pump site $X$ also reaches its maximum
($\sim$65\%).

It is evident from Fig.~2b that the occupation of the pump site $X$ is accompanied by the monotonic population of the the protonable
catalytic site $Z$, thus lowering the energy of the redox site $B$ from its initial level, $\varepsilon_B = -180$ meV, to the final
value of the order of $-820$ meV (see also Fig.~3). The population of the $B$-site, $ n_B $, closely follows (with a small delay)
the population $N_Z$ of the proton catalytic site $Z$ (see Figs.~2a and 2b). It can be seen from Fig.~2b that in $\sim$300
microseconds the pumped proton moves from the site $X$ to the transient site $C$, placed between $X$ and the P-side of the membrane,
and after that to the positive side of the membrane.

In the third phase ($ t\sim 1$~ms), the substrate (chemical) proton (Fig.~2b) occupies the catalytic site $Z$, $ N_Z > 0.8$. Then,
with a microsecond delay, the electron (Fig.~2a) is transferred, $ n_B \geq 0.8$, to the $B$-center (Cu$_B$), so that the heme $a$
is practically re-oxidized, $ n_L  \sim 0.02$. This stage is correlated with the $800~\mu$s phase mentioned in
Ref.~\cite{BelPNAS07}.

In the fourth phase ($t \sim 3$~ms), the pumped proton (Fig.~2c) moves to the positive site of the membrane, $ N_{\rm P} \simeq
0.95$, the substrate proton populates the site $Z$, $ N_Z = 1$, and the electron is almost completely transferred to the site $B$, $
n_B \simeq 1$. On average, about 1.3 protons are taken from the N-side of the membrane during the whole process.

The variations of the average electron energy, $E_{\rm el} = \langle H_{\rm el} \rangle,$ and the total energy of the system,
$$E_{\rm tot} = \langle H_0 \rangle + \mu_{\rm P} N_{\rm P} + \mu_{\rm N} N_{\rm N},$$ with time are shown in Fig.~3. Here $H_0$ is
the basic Hamiltonian of the system (\ref{H0}), $H_{\rm el}$ is the Hamiltonian of the electron component (\ref{Hel}), $N_{\rm P}$
and $N_{\rm N}$ are the average numbers of protons (\ref{Nsigma}) translocated to the P- or N-side of the membrane, respectively. At
the beginning, the electron has energy $$E_{\rm el}(0)\,\simeq~(\varepsilon_A - u_{AD})\, \simeq\, - 277 \;  {\rm meV},$$ and at the
end of the process its energy sinks to the level $$E_{\rm el}(5\  {\rm ms})\, \simeq \,(\varepsilon_B - u_{BZ} - u_{BD})\,  \simeq
\,- 828 \; {\rm meV}$$ with the total drop $\Delta E_{\rm el} \simeq 551$ meV, corresponding to the experimental value
\cite{Brez04}. The total energy of the system, $E_{\rm tot}$, shows a decrease of the order of $\Delta E_{\rm tot} \simeq 271$ meV,
which is less than the drop of electron energy since one proton  gains the energy during its pumping to the positive side of the
membrane.

\subsection{Pumping efficiency}
It follows from Fig.~2(c), that at  $t = t_B = 5$ ms, the average number of pumped protons, $ N_{\rm P}$, reaches its peak value,
which can be used as a definition \cite{KimPNAS07} of the pumping efficiency $\eta$: $\eta = N_{\rm P}(t_B)$. According to this
definition, the present model demonstrates an almost-perfect performance with an efficiency $\eta \simeq 0.95$ at $T~=~298$~K,
$\Delta \mu = 210$ meV, $V=150$ meV. This is comparable to the efficiency of cytochrome c oxidase \cite{Wik77,Brez04} pumping one
proton across the membrane per one electron consumed at the oxygen reduction site. We find that the definition of the efficiency
$\eta$ introduced above is not sensitive to the choice of the specific moment $t_B= 5$ ms, since the number of pumped protons,
$N_{\rm P}(t)$, does not decrease noticeably with time during the interval from 5 ms to more than 100~ms at the standard conditions.

In Fig.~4 we plot the pumping efficiency $\eta$ versus the transmembrane voltage $V$ at three different temperatures: $T = 150$ K
(blue dashed line), $T = 298$ K (green continuous curve), and $T = 450$ K (red dash-dotted line). We assume that the electrochemical
gradient $\Delta \mu$ varies in accordance to Eq.~(\ref{DMu}) where $\Delta pH = -1.$ At $T = 150$ K the pumping efficiency $\eta$
is almost constant at low voltages, $V < 150$ meV, with a subsequent drop at high voltages. The pump works better at room
temperatures, $T = 298$ K, and keeps the efficiency steady up to voltages $V \sim 200$ meV. Notice that in this case the efficiency
$\eta$, which is proportional to the average number of pumped protons, becomes negative at $V \geq 270$ meV. The performance of the
model is significantly deteriorated at high temperatures, $T = 450$ K, when the proton flow is reversed starting with the relatively
low voltage gradient, $ V \sim 180$ meV.

\section{Discussion}

The obtained time evolution of the electron and proton populations (see Fig.~2) features four experimentally observed phases of the
proton pumping process: the first ``$10~\mu$s" phase, when the electron is transferred from Cu$_A$ (site $A$) to heme a (site $L$);
the second ``$150~\mu$s" phase when the electron moves from heme a to heme $a_3$ (site $R$), and, with a microsecond delay, a proton
partially occupies the pump site $X$;  the third ``$1000~\mu$s" phase when the ``chemical" proton is transferred to the catalytic
sites $Z$ and, a slightly later, the electron is transferred to the ultimate electron acceptor Cu$_B$. In the fourth phase, at $t
\sim 3$ ms, the pre-pumped proton is released to the P-side of the membrane.

It should be emphasized that, contrary to the models proposed in Refs.\cite{BelPNAS07,WV07,SugStuch08}, this process cannot be
described as a sequence of transitions between clearly defined quasi-equilibrium states since many electron and proton transfers
occur in a very short time one after the other. The present theoretical model, which includes four redox sites (two copper centers
and two hemes) and four protonable sites, is able to explain the efficient performance ($\eta \sim 0.95$) of the real cytochrome c
oxidase \cite{Wik77} pumping almost one proton per one electron consumed against the electric potential difference, $V~\geq~150$
meV, and against the transmembrane electrochemical gradient, $\Delta \mu~\geq~210$ meV. We stress that all four kinetic phases
appear naturally in our model for a reasonable set of the system parameters without artificial inclusions of consequent transfer
processes.

The mechanism of the proton pumping analyzed above is based on the direct electrostatic interaction between the redox and protonable
sites, especially between the electron located on the site $R$ (heme $a_3$) and the proton located on the pump site $X$. The Coulomb
coupling between the redox site Cu$_B$ and the protonable catalytic site $Z$ plays a very important role as well. The proton to be
pumped is sequentially translocated to the P-side of the membrane from the sites $X$ and $C$. At the beginning of the process these
sites are empty since their energy levels are assumed to be higher than the energy levels of the proton source ($\mu_{\rm N}$ and
$E_D$) and the proton drain ($\mu_{\rm P}$). After the first ``$10~\mu$s" phase the energy level of the $L$-site is slightly ($\sim
55$ meV) lower than the energy level of the $R$-site. However, an interaction with the environment facilitates the slow electron
transfer to the site $R$. The population of the site $R$ with the electron is accompanied by the lowering of the $X$-site energy
level followed by the proton translocation from the site $D$ to the pump site $X$. Because of the strong $X$-$R$ electrostatic
attraction, the effective energy of the $R$-electron drops below the energy of the $L$-site, which results in the second
``$150~\mu$s" phase where the major part ($\sim 60$\%) of the electron density is concentrated on the site $R$, and the pump site
$X$ is partially ($\sim 65$\%) populated with a proton. The electron transfer to the site $R$ also leads to lowering the energy of
$Z$-site, thus inducing a monotonous population of the catalytic protonable site $Z$. No switch redirecting protons to the site $D$
or to the site $X$ (as proposed in Ref. \cite{WVH03}) is needed here because both of these sites can be populated from the site $D$.

It should also be emphasized that these three processes: the electron transfer to the $R$-site, the occupation of the pump site $X$,
and the translocation of a proton to the $Z$-site, are strongly correlated in time. The proton transfer to the $Z$-site digs a deep
potential well for the electron at the site $B$, and in the third ($\sim 1$ ms) phase the electron falls into this well. Afterwards,
the Coulomb attraction between the pre-pumped $X$-proton and the electron is almost compensated by the electrostatic repulsion
between $X$ and $Z$ protons, and the energy level of the $X$-proton returns to its original high value. The reverse translocation of
the $X$-proton to the site $D$ is strictly suppressed since now the energy difference between the sites $X$ and $D$ ($E_X - E_D \sim
380$ meV) significantly exceeds the reorganization energy $\lambda_{DX}$ as well as the temperature broadening, $2\sqrt{\lambda_{DX}
T}$, of the transition rates in Eq.~(\ref{KapAL}). However, the pre-pumped proton can easily move to the slightly ($\sim 25$~meV)
lower energy level $E_C$, and, after this, to the positive side of the membrane characterized by the even lower electrochemical
potential $\mu_{\rm P} = 105$ meV. Our model does not require any nonlinear gates \cite{WVH03} to prevent a proton leakage from the
positive to the negative side of the membrane.

\section{Conclusion}

We have analyzed a simple model describing the kinetics of the proton pumping process in cytochrome $c$ oxidase initiated by a
single-electron injection. Within our model, this electron is subsequently transferred along four sites electrostatically coupled to
four protonable sites. We have shown that the energy loss by this electron facilitates the proton transfer against the transmembrane
voltage from the negative to the positive sides of the membrane with the efficiency $\eta \sim 0.95.$ In contrast to previous
studies, we have not broken the electron and proton transfers into a series of transitions between the independent quasi-equilibrium
states but examined inseparable dynamics of the pumping process. We have derived the master equations of motion and solved them
numerically for a reasonable set of the system parameters. The obtained time evolution naturally encompasses \textit{all four}
experimentally observed kinetic phases.

\appendix

\section{Master equations}
The kinetics of charge transfer in the CcO enzyme can be described by a set of master equations \cite{KimPNAS07,Haken04,Ferreira06}.
For completeness we present here a derivation of these equations.
 We start from the formalism of second quantization
\cite{PumpPRE08,Wingr93,FlagPRE08}, even though in this paper we only discuss the classical results, with an examination of quantum
coherent effects to be performed in the future. Electrons, located on the redox sites $\alpha$ $(\alpha = A,L,R,B)$, are described
by the creation and annihilation Fermi operators $a_{\alpha}^+, a_{\alpha}$, and protons located on the protonable sites $\beta$
($\beta = D,X,C,Z$) are described by the creation and annihilation Fermi operators $b_{\beta}^+, b_{\beta}$. The spin degrees of
freedom are neglected; thus, each site can only be occupied by a single particle. The electron population of the $\alpha$-site,
$n_{\alpha}$, is expressed as $n_{\alpha} = a_{\alpha}^+a_{\alpha}$, and for a proton population $N_{\beta}$ on site $\beta$, we
have the relation: $N_{\beta} = b_{\beta}^+b_{\beta}$. Protons on the negative (N) and on the positive (P) side of the membrane
($\sigma$~=~N,P) are continuously distributed over the space of an additional ``quasi-wavenumber" parameter $q$ and characterized by
the creation and annihilation Fermi operators $d_{q\sigma}^+, d_{q\sigma }$ with the density operator $N_{q\sigma} =
d_{q\sigma}^+d_{q\sigma}$.

\subsection{Hamiltonian of the system}
The total Hamiltonian $H$ of the electron-proton system incorporates a basic term,
\begin{equation}
H_0 = \sum_{\alpha} \varepsilon_{\alpha} n_{\alpha} + \sum_{\beta} \varepsilon_{\beta} N_{\beta} + \sum_{\beta \beta'}
u_{\beta\beta'} N_{\beta}N_{\beta'} - \sum_{\alpha \beta} u_{\alpha\beta} n_{\alpha}N_{\beta}, \label{H0}
\end{equation}
 where the first and second terms describe the electron
($\alpha$) and proton ($\beta$) sites with energies $\varepsilon_{\alpha}$ and $\varepsilon_{\beta}$, respectively, and the third
and fourth terms are responsible for the Coulomb interaction of protons with each other and the electron, respectively. It should be
noted that in our single-electron model there is no inter-electron Coulomb interaction. We will also calculate the energy of the
electron component, which is determined by the Hamiltonian
\begin{equation}
H_{\rm el} = \sum_{\alpha} \varepsilon_{\alpha}n_{\alpha}  - \sum_{\alpha \beta} u_{\alpha\beta} N_{\beta} n_{\alpha}. \label{Hel}
\end{equation}
For protons in the N-side and P-side reservoirs we introduce the Hamiltonian
\begin{equation}
H_{\rm NP}~=~\sum_{q\sigma} \varepsilon_{q\sigma} N_{q\sigma}, \label{HNP}
\end{equation}
with the energy spectrum $\varepsilon_{q\sigma}$, whereas proton transitions between site $D$ and the N-side of the membrane, and
site $C$ and the P-side are given by the transfer Hamiltonian
\begin{equation}
H_{\rm tr} =  - \sum T_{q{\rm N}} d_{q{\rm N}}^+b_D  - \sum T_{q{\rm P}} d_{q{\rm P}}^+b_C + h.c., \label{Htr}
\end{equation}
characterized by the coefficients $T_{q{\rm N}}$ and $T_{q{\rm P}}$. The component
\begin{equation}
H_{\rm tun} = - \sum_{\alpha \alpha'} \Delta_{\alpha \alpha'}a_{\alpha}^+a_{\alpha'} - \sum_{\beta \beta'} \Delta_{\beta
\beta'}b_{\beta}^+b_{\beta'} \label{Htun}
\end{equation}
is responsible for electron tunneling between the pairs ($\alpha \alpha'$) of sites $A$-$L$, $L$-$R$, $R$-$B$, and for proton
transitions between the pairs ($\beta \beta'$) of sites $D$-$X$, $X$-$C$, and $D$-$Z$, with the corresponding amplitudes
$\Delta_{\alpha\alpha'}$ (for electrons) and $\Delta_{\beta\beta'}$ (for protons), where $\Delta_{\alpha\alpha'}^+ =
\Delta_{\alpha'\alpha}$ and $\Delta_{\beta\beta'}^+ = \Delta_{\beta'\beta}$.

Protons are delivered from a solution to the active site $D$ by the water-filled D-channel \cite{Gennis04,Branden06}. It was argued
\cite{Brez06,Nagle78,Wright06}  that the D-channel contains hydrogen-bonded chains of water molecules, which can convey protons via
the Grotthuss mechanism. In this case, the proton transfer is considered as a collective motion of a positive charge through the
chain, but not as a motion of an individual proton. According to another point of view (see Refs.
\cite{Olsson07,Olsson05,Kato06,BraunSand04}), the proper orientation of water molecules required for the Grotthuss mechanism is
characterized by a much smaller energetic penalty than the electrostatic barriers associated with a proton transfer through the
channel. Thus, the dominant contribution to the kinetic rate of the proton transport in proteins is provided by the electrostatic
energy. In the present work we model proton transitions (between the site $D$ and the N-side as well as between the site $C$ and the
P-side of the membrane) by the Hamiltonian $H_{\rm tr}$ (\ref{Htr}) with matrix elements that do not specify the transfer origin.

The transport of protons between the active sites $D$-$X$, $X$-$C$, and $D$-$Z$ are described  by phenomenological coefficients
$\Delta_{\beta\beta'}$ in the Hamiltonian (\ref{Htun}). To obtain kinetic rates for proton transitions between active sites in the
presence of an environment, we resort to the Marcus formulation of the problem. The relevant approach based on the empirical valence
bond method \cite{Warshel80} has been developed in Ref.~\cite{BraunSand04}. As shown in Refs.~\cite{Olsson06,Olsson07,Olsson05}, the
modified Marcus relations can be successfully applied for modelling the proton transfer steps in cytochrome $c$ oxidase.

\subsection{Environment}
To take into account the interaction of the electron-proton system with its environment, we introduce a term $H_{\rm env}$:
\begin{eqnarray}
H_{\rm env} = \sum_j \frac{p_j^2}{2m_j} + \frac{1}{2} \sum_j m_j\omega_j^2 \left(x_j - \sum_{\alpha} x_{j\alpha}n_{\alpha} -
\sum_{\beta} x_{j\beta}N_{\beta} - \sum_{\sigma} x_{j\sigma}N_{\sigma} \right)^2, \label{Henv}
\end{eqnarray}
where $N_{\sigma} = \sum_q N_{q\sigma}$ is the total number of protons in the $\sigma$-reservoir ($\sigma$ = N, P). The environment
is represented as a set of harmonic oscillators \cite{Garg85,Krish01} with coordinates $x_j$, momenta $p_j$, masses $m_j$, and
frequencies $\omega_j$. The shifts $x_{j\alpha}$, $x_{j\beta}$, and $x_{j\sigma}$ define coupling strengths of electrons and protons
to the environment. The total Hamiltonian $H$ is the sum of all above-mentioned components:
\begin{equation}
H = H_0 + H_{\rm NP} + H_{\rm tr} + H_{\rm tun} + H_{\rm env}. \label{Htot}
\end{equation}

With the unitary transformation,
\begin{equation}
U = \exp\left\{ - i \sum_j p_j \left( \sum_{\alpha} x_{j\alpha}n_{\alpha} + \sum_{\beta}x_{j\beta}N_{\beta} + \sum_{\sigma}
x_{j\sigma}N_{\sigma} \right) \right\} \label{UT}
\end{equation}
the total Hamiltonian, $H' = U^+HU,$ can be transformed to the form
\begin{eqnarray}
H' &=& H_0 + \sum_{q\sigma} \varepsilon_{q\sigma} N_{q\sigma} + \sum_j \left( \frac{p_j^2}{2m_j} + \frac{m_j\omega_j^2 x_j^2}{2}\right) \nonumber\\
&-& \sum_{\alpha \neq \alpha'} Q_{\alpha\alpha'} a_{\alpha}^+a_{\alpha'} - \sum_{\beta \neq \beta'} Q_{\beta
\beta'}b_{\beta}^+b_{\beta'}  \nonumber\\
&-& \sum T_{q{\rm N}} d_{q{\rm N}}^+b_D - \sum T_{q{\rm N}}^* b_D^+ d_{q{\rm N}} - \sum  T_{q{\rm P}} d_{q{\rm P}}^+b_C - \sum
T_{q{\rm P}}^* b_C^+ d_{q{\rm P}}, \label{Hnew}
\end{eqnarray}
where the operators,
\begin{eqnarray}
Q_{\alpha\alpha'} = Q_{\alpha'\alpha}^+ = \Delta_{\alpha \alpha'} \exp\{ i \sum_j p_j(x_{j\alpha} - x_{j\alpha'})\}, \nonumber\\
Q_{\beta\beta'} = Q_{\beta'\beta}^+ = \Delta_{\beta \beta'}\exp\{ i \sum_j p_j(x_{j\beta} - x_{j\beta'})\}, \label{QQ}
\end{eqnarray}
describe the effect of the environment on the electron and proton transitions. The protonable site $C$ is located near the P-side of
the membrane, and the site $D$ is tightly coupled to the N-side by the D-channel. It is reasonable to assume, therefore, that
$C$-to-P and N-to-$D$ proton transitions have a negligible effect on the equilibrium position of the $j$-oscillator of the
environment: $ x_{jC} = x_{j{\rm P}}, x_{jD} = x_{j{\rm N}},$ so that the corresponding phase factors in Eq.~(\ref{QQ}) related to
the Hamiltonian $H_{\rm tun}' = U^+H_{\rm tun}U$ and to the total Hamiltonian (\ref{Hnew}) can be omitted.

\subsection{Basis states and eigenenergies}
To quantitatively analyze  the system with a single electron and with up to four protons we introduce a basis of 64 eigenstates  of
the Hamiltonian $H_0$: $ |1\rangle = a_A^+|0\rangle, |2\rangle = a_A^+b_D^+|0\rangle,|3\rangle = a_A^+b_X^+|0\rangle, |4\rangle =
a_A^+b_C^+|0\rangle, |5\rangle = a_A^+b_Z^+|0\rangle, |6\rangle = a_A^+b_D^+b_X^+|0\rangle,\ldots ,|64\rangle =
a_B^+b_D^+b_X^+b_C^+b_Z^+|0\rangle.$ Here $|0\rangle$ is the vacuum state of the system with no electrons and no protons,
$|1\rangle$ is the state with an electron on site $A$, $|2\rangle$ is the state with an electron on site $A$ and a proton on site
$D$, $|3\rangle$ has one electron on site $A$ and a proton on site $X$, $|4\rangle$ describes the state with an electron on site $A$
and a proton on site $C$, and so on. Finally, $|64\rangle$ is the state with a single electron on site $B$ and with one proton on
each site $D,X,C,$ and $Z$ (i.e., a total of four protons). The state $|1\rangle$ has the eigenenergy $E_1 = \varepsilon_A$, the
state $|2\rangle$ has the energy $E_2 = \varepsilon_A + E_D - u_{AD},$ and the last state $|64\rangle$, fully loaded with four
protons, has the energy
$$
E_{64} = \varepsilon_B + \sum_{\beta=D}^{\beta=Z} ( \varepsilon_{\beta} - u_{B\beta} ) +  u_{DX} + u_{DC} + u_{DZ} + u_{XC} + u_{XZ}
+ u_{CZ}.$$ The Hamiltonian $H_0$ (\ref{H0}) is diagonal in the new basis:
\begin{equation}
H_0 = \sum_{\mu =1}^{64} E_{\mu} |\mu\rangle\langle \mu|, \label{H0mu}.
\end{equation}
Other operators may have a non-diagonal form in the new basis, for example,
\begin{eqnarray}
a_{\alpha}^+a_{\alpha'} = \sum_{\mu\nu} (a_{\alpha}^+a_{\alpha'})_{\mu\nu} \rho_{\mu\nu}, \nonumber\\
b_{\beta}^+b_{\beta'} = \sum_{\mu\nu} (b_{\beta}^+b_{\beta'})_{\mu\nu} \rho_{\mu\nu}, \nonumber\\b_{\beta} = \sum_{\mu\nu}
b_{\beta;\mu\nu} \rho_{\mu\nu},  \label{ab}
\end{eqnarray}
where
\begin{equation}
\rho_{\mu\nu} = |\mu\rangle\langle \nu|. \label{rho}
\end{equation}
Here indices $\mu$ and $\nu$ sweep all integers from 1 to 64.

\subsection{Electron and proton transitions}
In addition to the diagonal parts $H_0$ and $H_{\rm NP}$, the total Hamiltonian of the system $H$ contains the term responsible for
the proton transitions between the N-side of the membrane and the site $D$, and between the P-side and the site $C$:
\begin{equation}
H_{\rm tr} = - \sum ( T_{q{\rm N}}b_{D;\mu\nu} d_{q{\rm N}}^+ + T_{q{\rm P}}b_{C;\mu\nu} d_{q{\rm P}}^+ ) \rho_{\mu\nu} + h.c.
\label{HtunNew}
\end{equation}
as well as the off-diagonal term $H_{\rm tun}$ describing the tunneling of electrons and the transfer of protons between the active
sites,
\begin{equation}
H_{\rm tun} = - \sum_{\mu\nu} {\cal A}_{\mu\nu} \rho_{\mu\nu} - \sum_{\mu\nu}  \rho_{\nu\mu} {\cal A}_{\mu\nu}^+, \label{HtrNew}
\end{equation}
Here the operator $\cal{A}_{\mu\nu}$ is represented by a linear combination of the bath operators $Q_{AL},..,Q_{ZD}$ [see
Eqs.~(\ref{QQ})], multiplied by the non-diagonal ($\mu\neq\nu$) transition matrix elements
$(a_A^+a_L)_{\mu\nu},..,(b_Z^+b_D)_{\mu\nu}$:
\begin{eqnarray}
{\cal A}_{\mu\nu} &=&  Q_{AL} (a_A^+a_L)_{\mu\nu} +  Q_{LR} (a_L^+a_R)_{\mu\nu} +  Q_{RB} (a_R^+a_B)_{\mu\nu}  \nonumber\\
&+& Q_{DX} (b_D^+b_X)_{\mu\nu} + Q_{XC} (b_X^+b_C)_{\mu\nu} + Q_{DZ} (b_D^+b_Z)_{\mu\nu}. \label{Amunu}
\end{eqnarray}
It should also be noted that operators of the N and P proton reservoirs, $d_{q{\rm N}}$ and $d_{q{\rm P}},$ cannot be completely
expressed in terms of the basis operators $\rho_{\mu\nu}$.

\subsection{Derivation of the master equations}
A probability $\langle \rho_{\mu}\rangle$  to find the electron-proton system in the state $|\mu\rangle$ is determined by the
diagonal operator $\rho_{\mu} = |\mu\rangle\langle \mu|$ averaged over the states of the environment and over the distributions of
protons on both, N and P, sides of the membrane. The time evolution of the operator $\rho_{\mu}$ is governed by the Heisenberg
equation
\begin{eqnarray}
i \dot{\rho}_{\mu} = [\rho_{\mu}, H_{\rm tr}]_- - \sum_{\nu}\{ {\cal A}_{\mu\nu}\rho_{\mu\nu} - {\cal A}_{\nu\mu}\rho_{\nu\mu}\} +
\sum_{\nu}\{ {\cal A}_{\mu\nu}\rho_{\mu\nu} - {\cal A}_{\nu\mu}\rho_{\nu\mu}\}^+. \label{EqRho}
\end{eqnarray}
To derive a master equation for the probabilities $\langle \rho_{\mu}\rangle$, we have to average Eq.~(\ref{EqRho}) and calculate
the correlation functions $\langle {\cal A}_{\mu\nu}\rho_{\mu\nu}\rangle$  of the environment operators ${\cal A}_{\mu\nu}$
(\ref{Amunu}) and the operators of the system $\rho_{\mu\nu}$. The transition coefficients, $\Delta_{\alpha\alpha'},
\Delta_{\beta\beta'}$ and $T_{q\sigma}$ are supposed to be much smaller than the energy scales given by the basis spectrum
$E_{\mu}\; (\mu=1,\ldots,64).$ This means that the effective interactions with the N and P proton reservoirs [see
Eq.~(\ref{HtunNew})] and with the bath of oscillators (see Eqs.~(\ref{HtrNew}),(\ref{Amunu})) can be treated as a perturbation. In
the framework of the theory of open quantum systems proposed in Ref.~\cite{ES81} the correlation function $\langle {\cal
A}_{\mu\nu}\rho_{\mu\nu}\rangle$ (with $\mu \neq \nu$, no summation over $\mu$ and $\nu$)  can be written in the form
\begin{eqnarray}
\langle {\cal A}_{\mu\nu}(t)\rho_{\mu\nu}(t)\rangle &=&  \langle {\cal A}_{\mu\nu}^{(0)}(t)\rangle \langle \rho_{\mu\nu}(t)\rangle \nonumber\\
&+& \int dt_1 \langle {\cal A}_{\mu\nu}^{(0)}(t), {\cal A}_{\mu'\nu'}^{(0)+}(t_1)\rangle \langle i [ \rho_{\mu\nu}(t),\rho_{\nu'\mu'}(t_1)
]_-\rangle \theta(t-t_1) \nonumber\\ &+& \int dt_1 \langle i [ {\cal A}_{\mu\nu}^{(0)}(t), {\cal A}_{\mu'\nu'}^{(0)+}(t_1) ]_-\rangle \langle
\rho_{\nu'\mu'}(t_1) \rho_{\mu\nu}(t) \rangle \theta(t-t_1). \label{Furutsu}
\end{eqnarray}
Here ${\cal A}_{\mu\nu}^{(0)}(t)$ is a variable of the free environment (with no coupling to the electron-proton system), and
$\theta(t-t_1)$ is the Heaviside unit step function. We introduce the following notations for a cumulant function of two operators
${\cal A}(t)$ and ${\cal B}(t)$:
$$\langle {\cal A}(t), {\cal B}(t')\rangle = \langle {\cal A}(t) {\cal B}(t')\rangle - \langle {\cal A}(t)\rangle \langle {\cal B}(t')\rangle, $$ and
for a commutator: $$[{\cal A}(t), {\cal B}(t')]_{-} =  {\cal A}(t) {\cal B}(t') -   {\cal B}(t') {\cal A}(t).$$ In
Eq.~(\ref{Furutsu}) we take into account the backaction of the bath  in a contrast to the approach of Ref.~\cite{PumpPRE08} where
this backaction is not included into consideration. Due to significant decoherence effects, off-diagonal elements of the density
matrix, $\langle \rho_{\mu\nu}(t)\rangle$, disappear very fast. Because of this, the first term in the r.h.s.~of Eq.~(\ref{Furutsu})
can be neglected despite the non-zero value of the average unperturbed operator $\langle {\cal A}_{\mu\nu}^{(0)}(t)\rangle$. The
times $t$ and $t_1$ involved in the integrands of Eq.~(\ref{Furutsu}) are separated by the correlation time $\tau_c$ of the
correlators, which are similar to the function $\langle Q_{AL}(t),Q_{AL}(t_1)\rangle$. The timescale $\tau_c$ is determined by the
reorganization energy $\lambda_{AL}$ and temperature $T$: $\tau_c \sim \hbar /\sqrt{\lambda_{AL} T}$ (see
Refs.~\cite{Marcus56,Krish01} and Eq.~(\ref{QQSimple}) below). We assume that transitions between the active sites have a negligible
effect on the time evolution of the operator $\rho_{\mu\nu}$ between the times  $t$ and $t_1$, which are separated by the
correlation time $\tau_c$. Thus, the correlation functions and commutators of the operators $\rho_{\mu\nu}(t)$ and
$\rho_{\nu\mu}(t_1)$ can be calculated using free-evolving functions:
$$\rho_{\mu\nu}(t) = \rho_{\mu\nu}(t_1) \exp\{i\omega_{\mu\nu}(t-t_1)\},$$ where $\omega_{\mu\nu} = E_{\mu}- E_{\nu}.$ For the correlator
(\ref{Furutsu}) we obtain the formula
\begin{eqnarray}
\langle {\cal A}_{\mu\nu}(t)\rho_{\mu\nu}(t)\rangle = i \int dt_1 e^{i\omega_{\mu\nu}(t-t_1)} \theta(t-t_1) \times \nonumber\\
\{ \langle {\cal A}_{\mu\nu}^{(0)}(t), {\cal A}_{\mu\nu}^{(0)+}(t_1)\rangle \langle \rho_{\mu}(t) \rangle - \langle {\cal
A}_{\mu\nu}^{(0)+}(t_1), {\cal A}_{\mu\nu}^{(0)}(t)\rangle \langle \rho_{\nu}(t) \rangle \}. \label{FurNew}
\end{eqnarray}
With Eq.~(\ref{Amunu}),  we can express the cumulant $ \langle {\cal A}_{\mu\nu}^{(0)}(t), {\cal A}_{\mu\nu}^{(0)+}(t_1)\rangle $ in
terms of cumulant functions of the unperturbed bath operators $Q_{AL}^{(0)},\ldots,Q_{ZD}^{(0)}$:
\begin{eqnarray}
\langle {\cal A}_{\mu\nu}^{(0)}(t), {\cal A}_{\mu\nu}^{(0)+}(t_1)\rangle = |(a_A^+a_L)_{\mu\nu}|^2 \langle
Q_{AL}^{(0)}(t),Q_{AL}^{(0)+}(t_1)\rangle +  |(a_L^+a_R)_{\mu\nu}|^2 \langle Q_{LR}^{(0)}(t),Q_{LR}^{(0)+}(t_1)\rangle  \nonumber\\
+ |(a_R^+a_B)_{\mu\nu}|^2 \langle Q_{RB}^{(0)}(t),Q_{RB}^{(0)+}(t_1)\rangle + |(b_D^+b_X)_{\mu\nu}|^2 \langle
Q_{DX}^{(0)}(t),Q_{DX}^{(0)+}(t_1)\rangle  \nonumber\\ + |(b_X^+b_C)_{\mu\nu}|^2 \langle Q_{XC}^{(0)}(t),Q_{XC}^{(0)+}(t_1)\rangle +
|(b_D^+b_Z)_{\mu\nu}|^2 \langle Q_{DZ}^{(0)}(t),Q_{DZ}^{(0)+}(t_1)\rangle. \label{AACor}
\end{eqnarray}
The correlation function $\langle {\cal A}_{\mu\nu}^{(0)+}(t_1), {\cal A}_{\mu\nu}^{(0)}(t)\rangle$ has a similar form, with
cumulants $\langle Q_{AL}^{(0)}(t),Q_{AL}^{(0)+}(t_1)\rangle, \ldots ,$ being replaced by $\langle
Q_{AL}^{(0)+}(t_1),Q_{AL}^{(0)}(t)\rangle,\ldots, $ . Using the definitions (\ref{QQ}) of the bath operators we can calculate their
correlation functions. In particular,
\begin{eqnarray}
\langle Q_{AL}^{(0)}(t),Q_{AL}^{(0)+}(t_1) \rangle = |\Delta_{AL}|^2 \exp\{ -i {\cal W}^{(1)}_{AL}(t-t_1) \} \exp\{- {\cal
W}^{(2)}_{AL}(t-t_1) \}, \nonumber\\
\langle Q_{AL}^{(0)+}(t_1),Q_{AL}^{(0)}(t) \rangle = |\Delta_{AL}|^2 \exp\{ i {\cal W}^{(1)}_{AL}(t-t_1) \} \exp\{- {\cal
W}^{(2)}_{AL}(t-t_1) \}, \label{QQCor}
\end{eqnarray}
where
\begin{eqnarray}
{\cal W}^{(1)}_{AL}(\tau) = \sum_j\frac{m_j\omega_j}{2\hbar} (x_{jA}-x_{jL})^2 \sin \omega_j\tau, \nonumber\\
{\cal W}^{(2)}_{AL}(\tau) = \sum_j\frac{m_j\omega_j}{2\hbar} \coth\left(\frac{\hbar \omega_j}{2T}\right) (x_{jA}-x_{jL})^2 (1 -
\cos\omega_j\tau), \label{WW}
\end{eqnarray}
and $T$ is  the temperature of the environment ($k_B=1$). These expressions can be simplified in the high-temperature limit when the
thermal fluctuations are much faster ($\omega_j \tau \ll 1$) than the environment modes coupled to the charge transfer
\cite{Krish01}: ${\cal W}^{(1)}_{AL}(\tau) = \lambda_{AL} \tau,\; {\cal W}^{(2)}_{AL}(\tau) = \lambda_{AL} T \tau^2$, and,
correspondingly,
\begin{eqnarray}
\langle Q_{AL}^{(0)}(t),Q_{AL}^{(0)+}(t_1) \rangle = |\Delta_{AL}|^2 e^{-i\lambda_{AL}(t-t_1)} e^{-\lambda_{AL}T(t-t_1)^2}, \nonumber\\
\langle Q_{AL}^{(0)+}(t_1),Q_{AL}^{(0)}(t) \rangle = |\Delta_{AL}|^2 e^{i\lambda_{AL}(t-t_1)} e^{-\lambda_{AL}T(t-t_1)^2}.
\label{QQSimple}
\end{eqnarray}
We introduce here the reorganization energy,
\begin{equation}
\lambda_{AL} = \sum_j\frac{m_j\omega_j^2(x_{jA}-x_{jL})^2}{2}, \label{Lambda}
\end{equation}
corresponding to the electron transition from the site $A$ to the site $L$. Similar parameters can also be introduced for other
electron transitions: from $L$ to $R$, from $R$ to $B$, as well as for proton transitions between sites $D$ and $X$, $X$ and $C$,
and between $D$ and the catalytic site $Z$.

After a sequential substitution of Eqs.~(\ref{QQSimple}), (\ref{AACor}), (\ref{FurNew}) into the averaged equation (\ref{EqRho}), we
obtain the contribution of the inter-site transfers into the master equation
\begin{equation}
\langle \dot{\rho}_{\mu} \rangle =  \langle - i [ \rho_{\mu},H_{\rm tr} ]_-\rangle   + \sum_{\nu} \kappa_{\mu\nu} \langle
\rho_{\nu}\rangle - \sum_{\nu} \kappa_{\nu\mu} \langle \rho_{\mu}\rangle, \label{RhoM}
\end{equation}
where the combined rate $\kappa_{\mu\nu}$ contains contributions of all possible electron and proton transitions,
\begin{equation}
\kappa_{\mu\nu} =  (\kappa_{AL})_{\mu\nu} + (\kappa_{LR})_{\mu\nu} + (\kappa_{RB})_{\mu\nu} + (\kappa_{DX})_{\mu\nu}
+(\kappa_{XC})_{\mu\nu} + (\kappa_{DZ})_{\mu\nu}. \label{KapMunu}
\end{equation}
The rates corresponding to the specific electron transfers, $(\kappa_{AL})_{\mu\nu},(\kappa_{LR})_{\mu\nu},(\kappa_{RB})_{\mu\nu},$
and the rates related to the proton transfers, $(\kappa_{DX})_{\mu\nu},(\kappa_{XC})_{\mu\nu},(\kappa_{DZ})_{\mu\nu},$ are all
determined by the Marcus equations \cite{Marcus56,Krish01} with coefficients given by the appropriate transition matrices. In
particular,
\begin{eqnarray}
(\kappa_{AL})_{\mu\nu} = |\Delta_{AL}|^2 \sqrt{\frac{\pi}{\lambda_{AL}T} } \left( |(a_A^+a_L)_{\mu\nu}|^2 + |(a_A^+a_L)_{\nu\mu}|^2
\right) \exp\left[ - \frac{(E_{\mu} - E_{\nu} + \lambda_{AL} )^2}{4\lambda_{AL} T} \right], \label{KapAL}
\end{eqnarray}
\begin{eqnarray}
(\kappa_{DZ})_{\mu\nu} = |\Delta_{DZ}|^2 \sqrt{\frac{\pi}{\lambda_{DZ}T} } \left( |(b_D^+b_Z)_{\mu\nu}|^2 + |(b_D^+b_Z)_{\nu\mu}|^2
\right) \exp\left[ - \frac{(E_{\mu} - E_{\nu} + \lambda_{DZ} )^2}{4\lambda_{DZ} T} \right]. \label{KapDZ}
\end{eqnarray}
It should be noted that the ratio between the transposed rate coefficients is equal to the Boltzmann factor, as
$$\frac{(\kappa_{AL})_{\mu\nu}}{(\kappa_{AL})_{\nu\mu}} = \exp\left(-\; \frac{E_{\mu} - E_{\nu}}{T}\right),$$
which results in the Boltzmann distribution for the equilibrium density matrix of the system.

The contribution, $\langle - i [ \rho_{\mu},H_{\rm tr} ]_-\rangle$, of proton transitions between the site $D$ and the N-side of the
membrane and between the exit site $C$ and the P-side of the membrane  to the master equation (\ref{RhoM}) can be calculated with
the methods of quantum transport theory \cite{Wingr93,PumpPRE08,FlagPRE08}. The coupling to the proton reservoirs is described by
the relaxation matrix,
\begin{eqnarray}
\gamma_{\mu\nu} &=& \Gamma_{\rm N} \; \{ |b_{D;\mu\nu}|^2 [ 1 - F_{\rm N}(\omega_{\nu\mu}) ] + |b_{D;\nu\mu}|^2 F_{\rm
N}(\omega_{\mu\nu}) \} \nonumber\\ &+& \Gamma_{\rm P} \; \{ |b_{C;\mu\nu}|^2 [ 1 - F_{\rm P}(\omega_{\nu\mu}) ] + |b_{C;\nu\mu}|^2
F_{\rm P}(\omega_{\mu\nu}) \}, \label{gamma}
\end{eqnarray}
where the energy-independent coefficient,
\begin{equation}
\Gamma_{\sigma} = 2\pi \sum_q |T_{q\sigma}|^2 \delta (\omega - \varepsilon_{q\sigma}), \label{GamBig}
\end{equation}
determines the rate of a proton delivery to the $D$-site ($\sigma$ = N) or the rate of a proton removal from the $C$-site ($\sigma$
= P). We assume here that protons on the $\sigma$-side of the membrane are described by the Fermi distribution,
\begin{equation}
F_{\sigma}(\varepsilon_{q\sigma}) = \left[ \exp\left(\frac{\varepsilon_{q\sigma}- \mu_{\sigma}}{T} \right) + 1 \right]^{-1},
\label{Fermi}
\end{equation}
characterized by a chemical potential  $\mu_{\sigma}$.

As a result, we obtain the system of master equations for the probabilities $\langle \rho_{\mu}\rangle$, as
\begin{equation}
\langle \dot{\rho}_{\mu} \rangle =   \sum_{\nu} ( \kappa_{\mu\nu} + \gamma_{\mu \nu} ) \langle \rho_{\nu}\rangle - \sum_{\nu} (
\kappa_{\nu\mu} + \gamma_{\nu\mu} ) \langle \rho_{\mu}\rangle, \label{RhoMFin}
\end{equation}
where the inter-site rates $\kappa_{\mu\nu}$ are determined by Eqs. (\ref{KapMunu}),(\ref{KapAL}), and the relaxation matrix,
$\gamma_{\mu\nu}$, is given by Eq.(\ref{gamma}).

\subsection{Algebraic solution of the master equations}
Determination of the time-dependent solution of the master equations (\ref{RhoMFin}) can be reduced to a purely algebraic problem.
To accomplish this, we rewrite the equations (\ref{RhoMFin}) in the form
\begin{equation}
\langle \dot{\rho}_{\mu} \rangle = - \sum_{\nu} \Lambda_{\mu\nu} \langle \rho_{\nu} \rangle,
\end{equation}
with a total relaxation matrix $\Lambda_{\mu\nu}$, where $ \Lambda_{\mu\nu} = - (\kappa_{\mu\nu} + \gamma_{\mu\nu} )$ at $\mu\neq
\nu$, and $\Lambda_{\mu\mu} = \sum_{\nu} (\kappa_{\nu\mu} + \gamma_{\nu\mu} ).$ The vector $\langle \rho \rangle $ with the elements
$\langle \rho_{\mu}\rangle \; (\mu =1,..,64)$ can be represented as a sum of the steady-state part, $\rho^0$, and the time-dependent
deviation $\tilde{\rho}(t)$, as $\langle \rho \rangle = \rho^0 + \tilde{\rho}.$ Both the total probability vector $\langle
\rho\rangle $ and its steady-state value satisfy the normalization condition: $\sum_{\mu} \langle \rho_{\mu} \rangle = \sum_{\mu}
\rho_{\mu}^{0} = 1.$ The steady-state distribution can be found from the matrix equation $\Lambda \rho^0 = 0$, and for a
time-dependent part $\tilde{\rho}$ we have a rate equation in the form $(d/dt)\tilde{\rho} = - \Lambda \tilde{\rho}.$ Using the
unitary operator, ${\cal U}$, the matrix $\Lambda$ can be transformed to the diagonal form $\Lambda' = {\cal U}^{-1} \Lambda {\cal
U}$ with $\gamma_{\mu}'$ as the diagonal elements. This transformation should be accompanied by the transformation of the vector
$\tilde{\rho}$ as $ \tilde{\rho} = {\cal U} \rho'.$ Then, the vector $\rho'(t)$ obeys the diagonal equation with a simple solution
for its $\mu$-component: $\rho_{\mu}'(t) = e^{-\gamma_{\mu}' t} \rho_{\mu}'(0).$ Correspondingly, the time evolution of the
probability vector $\rho(t)$ from its initial value $\rho(0)$ is described by the formula
\begin{equation}
\langle\rho(t)\rangle = \rho^0 - {\cal S}(t) \rho^0 + {\cal S}(t) \rho(0), \label{RhoTime}
\end{equation}
where ${\cal S}(t) = {\cal U} {\cal Z}(t) {\cal U}^{-1}$, and ${\cal Z}(t)$ is the diagonal matrix with the elements ${\cal Z}_{\mu
\nu}(t) = \delta_{\mu\nu} e^{-\gamma_{\mu}' t}.$ It should be noted that ${\cal S}(0) = \hat{I},$ and ${\cal S}(\infty) = 0,$ where
$\hat{I}$ is the 64$\times$64 unit matrix.

\subsection{Proton current}
The time-dependent populations, $\langle n_{\alpha} \rangle$ and $\langle N_{\beta} \rangle$, of all redox and protonable sites in
the model are expressed in terms of the evolving probability distribution $\langle \rho(t)\rangle $. We recall that the index
$\alpha$ labels \ the redox sites $\alpha$ = $A$ (Cu$_A$), $L$ (heme $a$), $R$ (heme $a_3$), and $B$ (Cu$_B$). The index $\beta$
labels the protonable sites $\beta = D, X, C,$ and $Z$. Finally, the index $\sigma$ labels the two sides of the membrane $\sigma$ =
N, P. With the density matrix probability distributions, $\langle \rho_{\mu}(t)\rangle$,  over the states, $|\mu\rangle$, of the
system we can also find the proton flows from the N-side and P-side of the membrane into the system, $I_{\sigma} = (d/dt)\langle
N_{\sigma}\rangle,$ where $\langle N_{\sigma}\rangle $ is the total number of protons on the $\sigma$-side of the membrane: $\langle
N_{\sigma}\rangle = \sum_q \langle N_{q\sigma}\rangle $. Using techniques developed in quantum transport theory
\cite{Wingr93,PumpPRE08}, we obtain the formulas for the proton currents $I_{\rm N}$ and $I_{\rm P}$:
\begin{eqnarray}
I_{\rm N} &=& \Gamma_{\rm N} \sum_{\mu\nu} | b_{D;\mu\nu} |^2 \{ [ 1 - F_{\rm N}(\omega_{\nu\mu}) ] \langle \rho_{\nu} \rangle -
F_{\rm N}(\omega_{\nu\mu})
\langle \rho_{\mu} \rangle \}, \nonumber\\
I_{\rm P} &=& \Gamma_{\rm P} \sum_{\mu\nu} | b_{C;\mu\nu} |^2 \{ [ 1 - F_{\rm P}(\omega_{\nu\mu}) ] \langle \rho_{\nu} \rangle -
F_{\rm P}(\omega_{\nu\mu}) \langle \rho_{\mu} \rangle \}. \label{Current}
\end{eqnarray}
Note that these currents  depend on the time-dependent probability distribution $\langle \rho(t)\rangle $ and, accordingly, they
also vary with time. The total number of protons, $\langle N_{\sigma}\rangle$, transferred to the $\sigma$-side of the membrane
($\sigma$ = P,N) is calculated as the integral of the corresponding current:
\begin{equation}
\langle N_{\sigma}(t) \rangle = \int_0^t dt_1 \;I_{\sigma}(t_1). \label{Nsigma}
\end{equation}

\subsection{Proton-motive force}
The proton-motive force across the membrane can be defined as a difference of electrochemical potentials $\mu_{\rm P}$ and $\mu_{\rm
N}$ involved in the Fermi distributions (\ref{Fermi}) of the proton reservoirs: $\Delta \mu = \mu_{\rm P} - \mu_{\rm N}$. This
gradient includes the transmembrane concentration difference ($\Delta pH$) and the transmembrane voltage $V $:
\begin{equation}
\Delta \mu = V - 2.3 \,(RT/F)\times\Delta pH. \label{DMu}
\end{equation}
Here $R$ and $F$ are the gas and Faraday constant, respectively, and $T$ is the temperature (in degrees Kelvin, $k_B = 1$)
\cite{Alberts02,Nicholls92}. Both energy parameters, $\Delta\mu$ and $V$, are measured in meV. At the standard conditions ($T =
298~K$), the concentration gradient contributes about 60 meV per $\Delta pH$-unit. This results in the transmembrane voltage $V \sim
150 $ meV, provided that the total proton-motive force, $\Delta \mu $, is about 210 meV, and $\Delta pH = -1$ \cite{Alberts02}.

The transmembrane voltage, $V > 0,$ elevates the energies of protonable sites adjacent to the P-side and lowers the energies of the
proton sites located near the N-side \cite{KimPNAS07}. The electron sites are simultaneously experiencing the opposite effect, for
the same $V$. As a result the electron energy levels, $\varepsilon_{\alpha}$, and the proton energies, $\varepsilon_{\beta}$,
involved in the Hamiltonian $H_0$ (\ref{H0}) are shifted from their initial values, $\varepsilon_{\alpha}^{(0)}$ and
$\varepsilon_{\beta}^{(0)}$:
\begin{eqnarray}
\varepsilon_{\alpha} = \varepsilon_{\alpha}^{(0)} - V (x_{\alpha}/W), \nonumber\\
\varepsilon_{\beta} = \varepsilon_{\beta}^{(0)} + V (x_{\beta}/W), \label{EnergyV}
\end{eqnarray}
where $W$ is the membrane width. The positions of the redox and protonable sites, $x_{\alpha}$ and $x_{\beta}$, are counted here
from the middle of the membrane with the x-axis directed toward the P-side: $x_A\sim W/2,\  x_L\sim x_R \sim x_B \sim W/6,\ x_D\sim
0.1\  W,\  x_X\sim 0.3\ W,\ x_C=0.5\  W,\  x_Z=W/6 $ \cite{BelPNAS07,MedStuch05}.

\vspace{2cm}

\noindent \textbf{Acknowledgements.}

\noindent This work was supported in part by  the National Security Agency (NSA), Laboratory of Physical Science (LPS), Army
Research Office (ARO), National Science Foundation (NSF) grant No. 0726909, and JSPS-RFBR 06-02-91200. L.M. is partially supported
by the NSF NIRT, grant ECS-0609146.

\begin{figure}
\includegraphics[width=15.0cm]{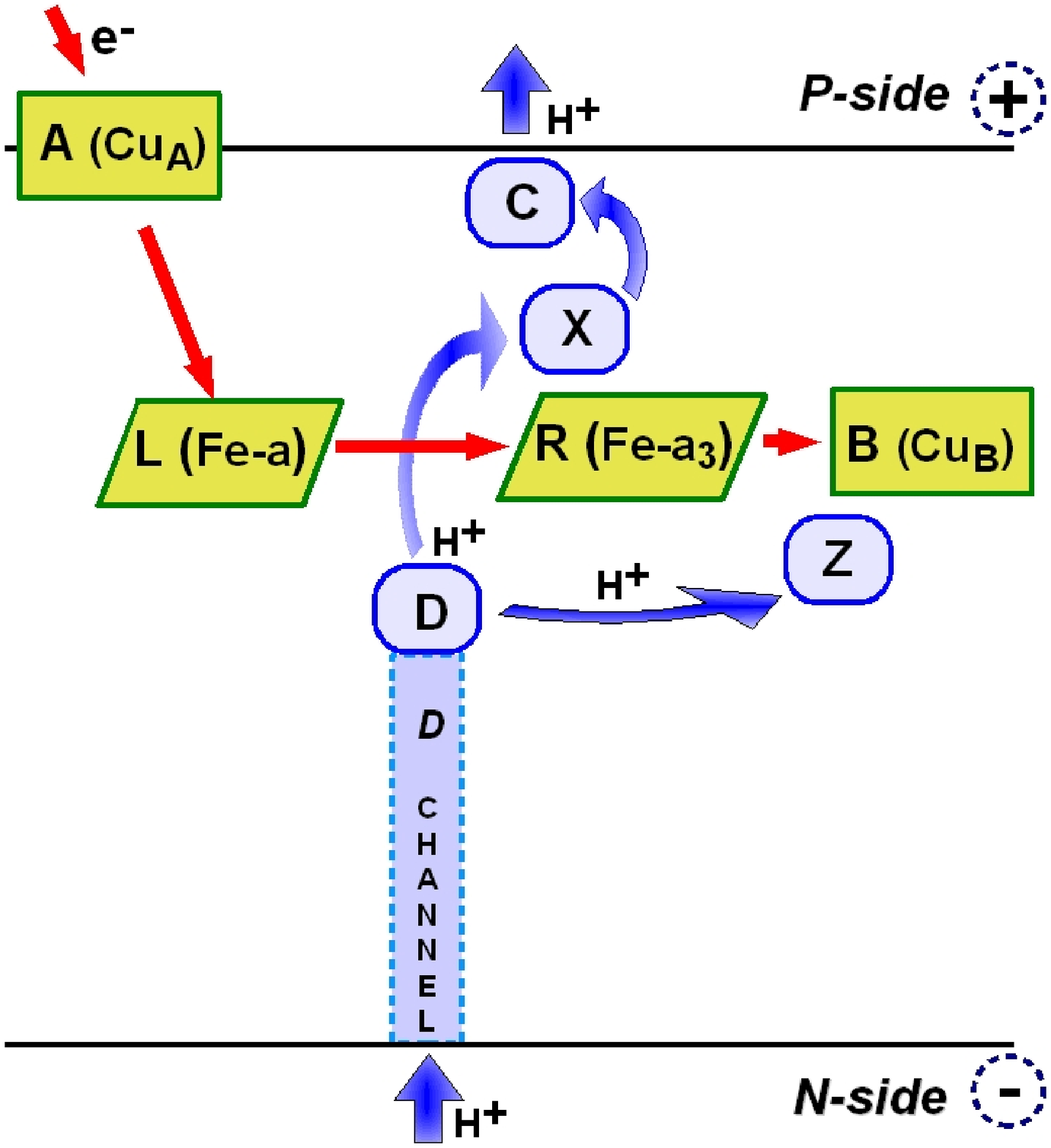}
\vspace*{-2cm} \caption{ (Color online) Schematic diagram of cytochrome $c$ oxidase. A single electron enters the enzyme at the site
$A$ and travels subsequently to sites $L$, $R$ and, finally, to site $B$. Protons, taken at the N-side of the membrane, move to site
$D$, which can donate protons both to the pump site $X$ and to the catalytic site $Z$. The pre-pumped proton is transferred from the
site $X$ to the P-side of the membrane via site $C$.}
\end{figure}

\begin{figure}
\includegraphics[width=16.5cm, height=11.0cm ]{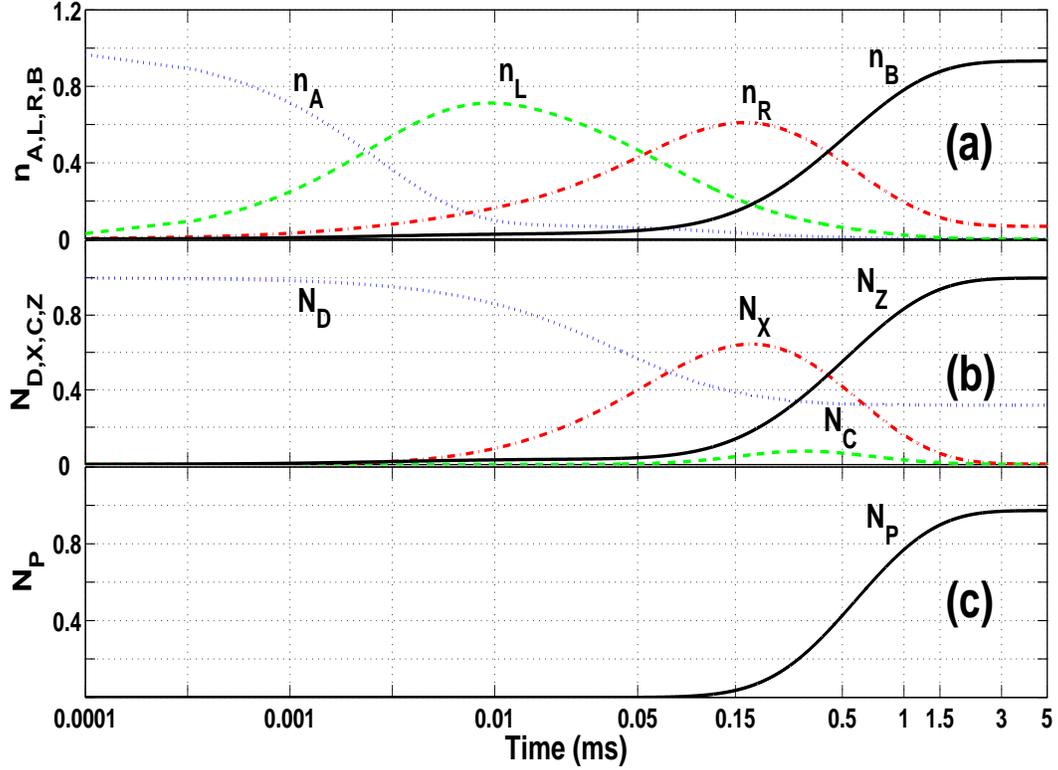}
\vspace*{-1.0cm} \caption{(Color online) Kinetics of electron and proton transfers in cytochrome $c$ oxidase for $\mu_{\rm P}~=~105$
meV, $\mu_{\rm N} = -105$ meV, $T = 298~K$, and $V = 150$ meV. The time axis (in ms) is shown in a logarithmic scale starting at $t
= 0.1~\mu$s. The process begins at $t = 0$, when a single electron populates the site $A$, and a single proton is located on the
site D. (a) Time dependence of the electron populations $n_A$ (blue dotted line), $n_L$ (green dashed line), $n_R$ (red dash-dotted
line), and $n_B$ (black continuous line). (b) The proton populations, $N_D$ (blue dotted line), $N_X$ (red dash-dotted line), $N_C$
(green dashed line), and $N_Z$ (black continuous line), versus time. (c) The number of pumped protons, $N_{\rm P}$, as a function of
time. The first phase of the process, where the electron moves from site $A$ to site $L$, corresponds to the maximum of the
population $n_L$ at the moment $t \approx 10\ \mu$s. In the second phase, both the electron population of the $R$-site, $n_R$, and
the proton population of the $X$-site, $N_X$, peak at $t \approx 150~\mu$s. The third phase of the process is marked by the
significant population of the proton site $Z$ and the electron site $B$  at $t \sim$ 1 ms. In this phase the site $X$ is
depopulated, and the pre-pumped proton is partially transferred to the P-side of the membrane, $N_{\rm P} \sim 0.8$. In the final
phase, at $t \sim$ 3 ms, the electron site $B$ and the catalytic proton site $Z$ are  occupied, $n_B \simeq 1,$ $N_Z$ =1, and about
one proton ($N_{\rm P} \sim 0.95$) is translocated to the P-side of the membrane. }
\end{figure}

\begin{figure}
\includegraphics[width=16.0cm]{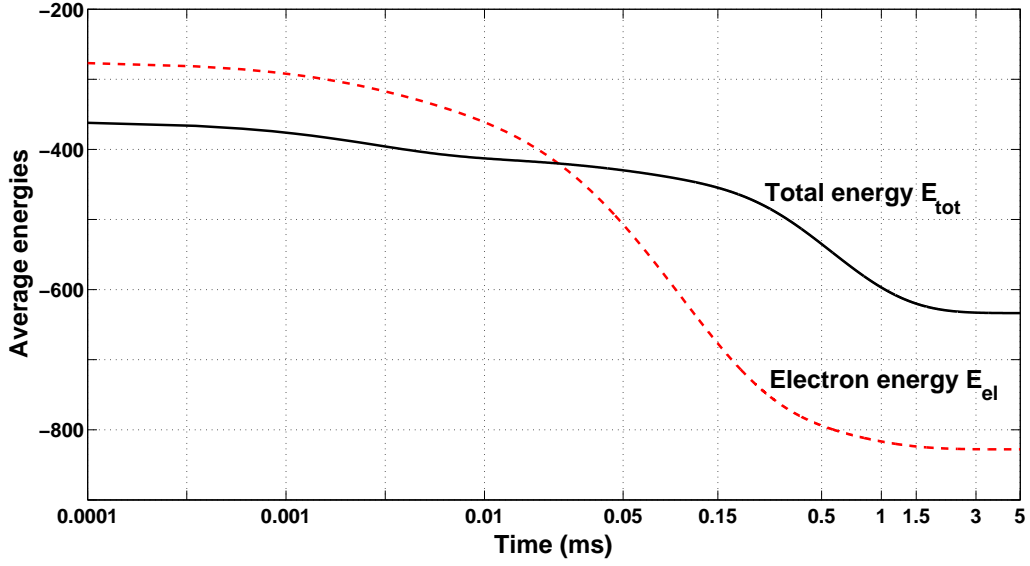}
\vspace*{0cm} \caption{(Color online) The total energy of the system, $E_{\rm tot}$ (black continuous line), and the average energy
of the electron, $E_{\rm el}$, as functions of time (in ms, logarithmic scale). The electron energy is varied in the range from
$E_{\rm el} = -277$~meV at $t\simeq 0$, to the value $E_{\rm el} = -828$~meV at $t = t_B = 5$~ms. The whole electron-proton system
dissipates less energy, $\Delta E_{\rm tot} \simeq 270$~meV, than its electron component, which loses about 550 meV during the
pumping process,  indicating the energy transfer to the proton subsystem. }
\end{figure}

\begin{figure}
\includegraphics[width=16.0cm]{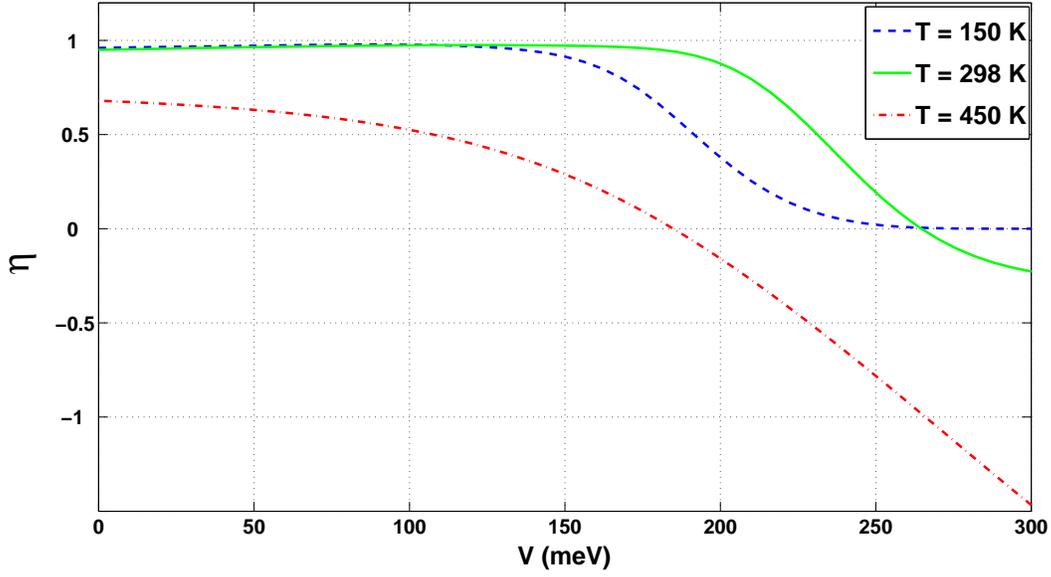}
\vspace*{0cm} \caption{(Color online) The efficiency of the pump, $\eta = N_{\rm P}(t_B),$ where $t_B = 5$ ms, as a function of the
transmembrane voltage $V$ at the temperatures $T = 150$ K (blue dashed line); $T = 298$ K (green continuous line), and $T = 450$ K
(red dash-dotted line). For the physiological range of transmembrane voltages, 150 meV $< V <$ 200 meV, the pump operates with a
maximum efficiency at room temperatures ($T = 298$ K). At high temperatures and high enough transmembrane voltages, the efficiency
$\eta$ takes negative values, suggesting that at these conditions the protons flow back, from the positive to the negative side of
the membrane.}
\end{figure}

\end{document}